\documentclass[10pt]{article}
\usepackage{fullpage}
\usepackage{graphicx}
\usepackage{color}
\usepackage{amssymb}
\usepackage{amsmath}
\usepackage{url}

\newcommand{\vir}[1]{``#1"}
\newcommand{\ignore}[1]{}

\title{The Chromatin Organization of an Eukaryotic Genome : Sequence Specific+ Statistical=Combinatorial \break (Extended Abstract)\footnote{Work presented at the 8th SIBBM Seminar (Annual Conference Meeting of the Italian Biophysics  and Molecular Biology Society)- May 24-26 2012, Palermo, Italy}}
\author{
Davide Corona\thanks{Dulbecco Telethon Institute c/o
Universit\`{a} di Palermo, Dipartimento di Biologia Cellulare e dello Sviluppo, Palermo, Italy;
davide.corona@unipa.it}
\and  Valeria Di Benedetto\thanks{Universit\`{a} di Palermo, Dipartimento  di
Matematica ed Informatica, Palermo, Italy;
valeria.dibenedetto@gmail.com}
\and  Raffaele Giancarlo\thanks{Universit\`{a} di Palermo, Dipartimento  di
Matematica ed Informatica, Palermo, Italy;
raffaele@math.unipa.it}
\and Filippo Utro\thanks{Computational Genomics Group, IBM T.J. Watson Research Center, Yorktown Heights, USA;
futro@us.ibm.com}
}

\begin{document}

\maketitle


\abstract{Nucleosome organization in eukaryotic genomes has a deep impact on gene function. Although progress has been recently made in the identification of various concurring factors influencing nucleosome positioning, it is still unclear whether nucleosome positions are sequence dictated or determined by a random process. It has been postulated for a  long time that, in the proximity of TSS, a ‘barrier’ determines the position of the +1 nucleosome and then geometric constraints alter the random positioning process determining nucleosomal phasing. Such a pattern fades out as one moves away from the barrier to become again a random positioning process. Although this statistical model is widely accepted, the molecular nature of the ‘barrier’ is still unknown. Moreover, we are far from the identification of a set of sequence rules able: to account for the genome-wide nucleosome organization; to explain the nature of the barriers on which the statistical mechanism hinges; to allow for a smooth transition from sequence-dictated to statistical positioning and back. Here we show that sequence complexity, quantified via various methods, can be the rule able to at least partially account for all the above. In particular, we have conducted our analyses on four high resolution nucleosomal maps of the model eukaryotes S.cerevisiae, C. elegans and D.melanogaster, and found that nucleosome depleted regions can be well distinguished from nucleosome enriched regions by sequence complexity measures. In particular, the depleted regions are less complex than the enriched ones. Moreover, around TSS, complexity measures alone are in striking agreement with in vivo nucleosome occupancy, in particular precisely indicating the positions of the +1 and -1 nucleosomes. Those findings indicate that the intrinsic richness of subsequences within sequences plays a role  in nucleosomal formation in genomes, and that sequence complexity constitutes the molecular nature of nucleosome ‘barrier’. }

\section{Background}
It is well known that chromatin organization in  Eukaryotic  genomes has a deep impact on gene regulation and function, e.g., \cite{FelGro2003}.
Therefore, it comes as no surprise that a substantial amount of research has been devoted to this fascinating topic, with particular focus on the identification of nucleosome positions within the genomes of model organisms and of mechanisms influencing  that positioning. Although quite a bit of progress has been made on the identification of various concurring factors that influence  nucleosome positioning \cite{SW09}, an answer to the  question raised by Kornberg in 1981 \cite{Korn81} as to which extend  nuclesome positions are \vir{sequence dictated} or determined by a random process has remained elusive.
As a matter of fact, it has been the object of intense debate, in particular in view of a result by Widom et al. \cite{SW06} claiming that there exists a genomic  code for nucleosome  positioning. In mathematical terms, a code is a very specific and constrained object, even when it is degenarate as the genetic code. Therefore, it seems that a {\em{verbatim}}  use of the term code  in the context of chromatin studies would be quite misleading and actually very easy to challenge, e.g.,  \cite{Val08,Zhang09}. Indeed, the focus has shifted from \emph{sequence dictating} to \emph{sequence  influencing}  chromatin organization, a finding that is much less amenable to challenge and dismissal \cite{Fields08}. In contrast to the \vir{sequence dictating-influencing} debate just outlined, the statistical positioning mechanism has had very little challenge. It is worth recalling that it is based on two main ingredients: the  existence of a barrier and of a statistical law governing the positioning of the nucleosomes \cite{KornStryer}. Generalizing the just mentioned earlier results by Kornberg and Stryer, M\"{o}bius and Gerland \cite{MG10} have shown that such a mechanism can be described and quantified by the Tonks model of statistical physics. In particular, in the proximity of TSS \cite{MP08,MG10}, the barrier determines the position of the +1 nucleosome and then  geometric constraints alter the random positioning process determining the well known nucleosomal pattern in DNA. Such a pattern fades out as one moves away from the barrier to become again a random positioning process. An analogous behaviour is followed by the -1 nucleosome.

\section{Statement of Results}
All of the above studies leave open several questions. Indeed, given the poor results in finding a consistent and concise set of \vir{sequence-rules} , e.g., \cite{Val08,Zhang09}, that explain how the sequence influences nucleosome positioning, it is a challenging problem to show that such a set of rules exists. Analogously,
although  the statistical model is widely accepted \cite{SW09,Zhang09}, the molecular nature of the barrier is unknown \cite{Zhang09}. Moreover, the interplay between sequence-dictated vs statistical positioning has been hardly explored, although it is to be expected that   the true cellular state is probably a combination of both machanisms \cite{JP10}. A study by Mavrich et al.  \cite{MP08}, performed on DNA regions around 4799 TSS in Yeast,  addresses in part this latter problem indicating  that the sequence dictates the  positioning of the so-called +1 and -1 nucleosomes, leaving the rest to the statistical positioning mechanism. In that study,  an effort is also made to identify some of the compositional properties of the nucleosome-free region (NFR)   responsible for the formation of the barriers. However, it is also pointed out that the identified sequence biases, mostly for dinucleotides,   may be special cases of more general, and yet unknown,   properties of the genomic sequences up and downstram of a TSS. In a nutshell,  we are far from the identification  of  a concise set of sequence rules able: (a) to account for the genome-wide nucleosome organization  in a genome; (b) to explaing the nature  of the barriers on which the statistical mechanism hinges; (c) to  allow for a smooth transition from sequence-dictated to statistical positioning and back.

As said before, a code for nucleosome positioning seems to be too stringent. Fortunately, there are other intrinsic properties of sequences, in particular the ones measured by complexity measures, that may play a role in influencing nucleosomal organization in a eukaryotic genome. It is worth recalling that complexity measures usually quantify the \vir{intrinsic richness} of distinct subsequences within a given sequence. Here we study whether sequence complexity, quantified via various methods, can be the rule able to at least partially account for all (a)-(c) above. Towards this end, we  have the following results.

\begin{itemize}

\item[(1)] We have conducted extensive studies on four  high resolution nucleosomal maps of three model orgamisms, i.e.,  Yeast, \emph{C. Elegans} and \emph{Drosophila Melanogaster} \cite{Fields08,MP08b,Zhang09}, to obtain the following results.     Nucleosome  depleted regions in each  map can be well distinguished from  nuclesome enriched regions in each  map by sequence complexity measures. In particular, the depleted regions are less complex than the enriched ones. Such a finding indicates that the intrinsic richness of subsequences within sequences plays a role in influencing nucleosomal formation in genomes, addressing point (a) above.

\item[(2)] We have applied our methodology to the same TSS dataset of Mavrich  et al. \cite{MP08}, and also accounted for the additional insights by M\"{o}bius and Gerland \cite{MG10}.
We find that the NFR is characterized by an area of lower complexity with respect to the two regions flanking it, with the TSS being placed in the proximity of  the absolute minimum of each complexity curve we have computed. The two barriers are towards the local maxima at the left and right of the TSS.
Particularly striking is the complexity curve obtained with linguistic complexity, based on a dictionary of words of length up to seventeen (therefore much larger than the one considered by Mavrich et al.). Indeed,
the positions of the +1 and -1 nucleosomes, respectively,  are in the proximity of the first maximum to the right and to  the left, respectively,  of the TSS on the complexity curve. In view of point (b) above,  such a finding suggests, at least as far as TSS are concerned, that the combinatorial properties of the NFR subsequence are strongly associated with the creation of the barriers, highlighting that its nature is at least in part combinatorial, i.e., dictated by intrinsic properties of sequences.

\item[(3)] Since the complexity of subsequences within  a sequence can be modulated, points (1) and (2) above suggest that, at least for TSS, such a modulation can smoothly accomodate for both sequence-dictated and statistical positioning, in a \emph{continuum} that requires no  other particular arrangement for switching between the two  \vir{states} of interest. Indeed, the low complexity area characterizing the NFR indicated where nucleosomes should not be, giving also an indication that the +1 and -1 nucleosomes must be placed towards the local maxima of the complexity curve at the end of the NFR. Such an interpretation sheds some light on point (c) above.

\end{itemize}

\bibliographystyle{plain}
\bibliography{nucleosome}

\end{document}